\documentclass[preprint,amsmath,amssymb]{revtex4-1}

\usepackage{graphicx}
\usepackage{dcolumn}
\usepackage{bm}

\begin{document}


\title{Orientation Determination in Single Particle X-ray Coherent Diffraction Imaging Experiments}

\author{O.M. Yefanov}
\email{oleksandr.yefanov@desy.de}
\altaffiliation[present address: ]{Center for Free-Electron Laser Science, Notkestra{\ss}e 85, D-22607 Hamburg, Germany.}
\affiliation{Deutsches Elektronen-Synchrotron (DESY), Notkestra{\ss}e 85, D-22607 Hamburg, Germany}%
\author{I.A. Vartanyants}%
\affiliation{Deutsches Elektronen-Synchrotron (DESY), Notkestra{\ss}e 85, D-22607 Hamburg, Germany}%
\affiliation{National Research Nuclear University "MEPhI", 115409 Moscow, Russia}%


\date{\today}

\begin{abstract}
Single particle diffraction imaging experiments at free-electron lasers (FEL) have a great potential for structure determination of reproducible biological specimens that can not be crystallized. One of the challenges in processing the data from such an experiment is to determine correct orientation of each diffraction pattern from samples randomly injected in the FEL beam. We propose an algorithm \cite{OrientReported} that can solve this problem and can be applied to samples from tens of nanometers to microns in size, measured with sub-nanometer resolution in the presence of noise. This is achieved by the simultaneous analysis of a large number of diffraction patterns corresponding to different orientations of the particles. The algorithms efficiency is demonstrated for two biological samples, an artificial protein structure without any symmetry and a virus with icosahedral symmetry. Both structures are few tens of nanometers in size and consist of more than 100 000 non-hydrogen atoms. More than 10 000 diffraction patterns with Poisson noise were simulated and analyzed for each structure. Our simulations indicate the possibility to achieve resolution of about 3.3 {\AA} at 3 {\AA} wavelength and incoming flux of 10$^{12}$ photons per pulse focused to 100$\times$100 nm$^2$.
\end{abstract}

\pacs{42.30.Rx, 42.30.Wb, 41.60.Cr}


\maketitle


\section{\label{sec:level1}Introduction}

The problem of solving the structure of individual biological specimens to high resolution is critical for many branches of modern life- and bio-science. Two widely used techniques for high resolution structure determination are x-ray crystallography and electron microscopy. X-ray crystallography can only be used for molecules that form crystals
\cite{ProteinCryst}, whereas transmission electron microscopy is limited to structures with a thickness well below one micron \cite{Beck2007}. Therefore the majority of samples must be sliced \cite{SampleEM}  and the minimum thickness of the slices limits the resolution of this method.

Single particle coherent diffraction imaging \cite{Neutze2000,Gaffney2009,Mancuso2010} is one of the promising new techniques for the investigation of biological samples to subnanometer resolution. It has become possible only recently due to the development of x-ray free-electron lasers \cite{Ackermann2007, Emma2010, Ishikawa2012, XFELTDR} , which produce ultra-short (less than 100 fs), coherent x-ray pulses with high intensity (more than $10^{12}$ photons in a single pulse). Short and intense pulses are required to overcome the radiation damage of biological particles during the pulse propagation \cite{Howells2009,Quiney2011,Lorenz2012} and to produce a high number of elastically scattered photons \cite{Neutze2000, Bergh2008}. The coherence of the incident beam is important for a successful reconstruction of the electron density of the sample \cite{Vartanyants2001, Vartanyants2003, Williams2007}. However, after the pulse propagation the particles will explode, and only one projection of the sample can be measured. This problem can be overcome by injecting particles one after another with random orientations and collecting a set of diffraction patterns \cite{Gaffney2009}. Each measured diffraction pattern corresponds then to an unknown particle orientation. A method to determine the orientation of the particle, corresponding to each  diffraction pattern, is the main subject of this paper. When the relative angular orientation of all diffraction patterns is determined the full three-dimensional (3D) intensity distribution in reciprocal space can be obtained. The structural information, or electron density of the sample is determined then by the phase retrieval \cite{Fienup1982,Marchesini2007}.

During the last few years there was a significant progress in the practical implementation of these ideas at hard x-ray FELs (see, for example, \cite{Seibert2011,Kassemeyer2012,Loh2012}).
There were few attempts to determine three-dimensional (3D) structure in single particle imaging experiments \cite{Loh2010}, however, the methods are still under development.
Several approaches have been proposed so far to find an unknown particle orientation in these experiments. One is based on the common arc algorithm \cite{Huldt2003} originally developed for electron microscopy \cite{VainshteinGoncharov1986,VanHeel1987}. This algorithm exploits the fact that all two-dimensional (2D) diffraction patterns of reproducible particles in random orientations represent sections by the Ewald sphere of the 3D intensity distribution in reciprocal space. As such, all diffraction patterns have one common point, the origin of reciprocal space, and intersect along common arcs. The intensities along these arcs must be equal, and using this information the relative orientation of all diffraction patterns can be determined. The main problem of this method is its demand for a high signal to noise ratio, which is difficult to satisfy even with the present high power FEL sources. It was suggested to overcome this limitation by an additional classification  step \cite{Huldt2003,Bortel2009}, in which diffraction patterns with similar particle orientations are averaged prior to orientation determination. This step improves the statistics of each averaged diffraction pattern, but at the same time reduces its contrast. As a result, the classification step decreases the achievable resolution and can produce artifacts in the final stage of electron density reconstruction. Another method is based on generative topographic mapping and neural networks \cite{Fung2009,Schwander2010}. This approach works well for a low signal to noise ratio but scales poorly with the number of resolution elements in terms of computational time and memory. The same is valid for a method based on an expectation maximization technique \cite{Elser2009}.

Here, we propose an orientation determination method based on an improved common arc algorithm \cite{OrientReported}. Instead of a classification step we perform a simultaneous analysis of common arcs between many diffraction patterns. To improve the quality of the orientation determination, a 3D angular refinement procedure is applied at the final step. This algorithm works well even with a low photon signal down to 0.05 photons per pixel for sampling rate of three at the edge of the detector. It scales linearly with the number of resolution elements and number of measured diffraction patterns. Memory requirements are relaxed because most of the data can be processed in parts. Finally, the algorithm is highly parallelizable since most of the analysis is done between pairs of diffraction patterns.

The paper is organized in the following way. In section two we describe our implementation of the common arc algorithm. Section three describes our approach to treat poor signal to noise ratio data as well as orientation refinement procedure. Tests of the proposed algorithm on simulated data from two different biological structures are presented in section four. The paper is completed by the conclusion section. The details of the algorithm implementation are presented in the Appendix.

\section{\label{sec:common}Common arc algorithm}

\begin{figure*}[p]
\centerline{\includegraphics[width=0.45\textwidth]{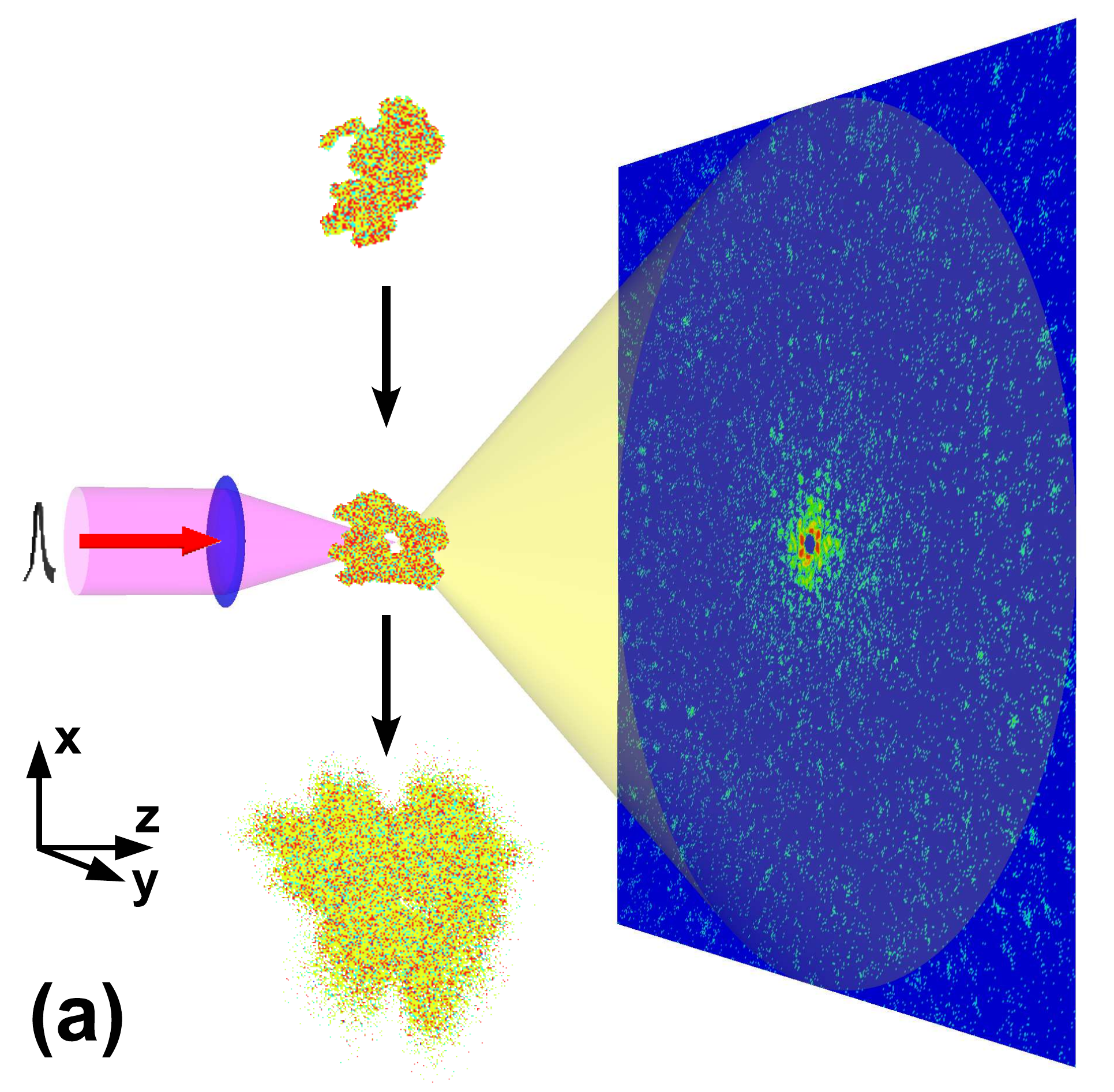} \hfil
\includegraphics[width=0.45\textwidth]{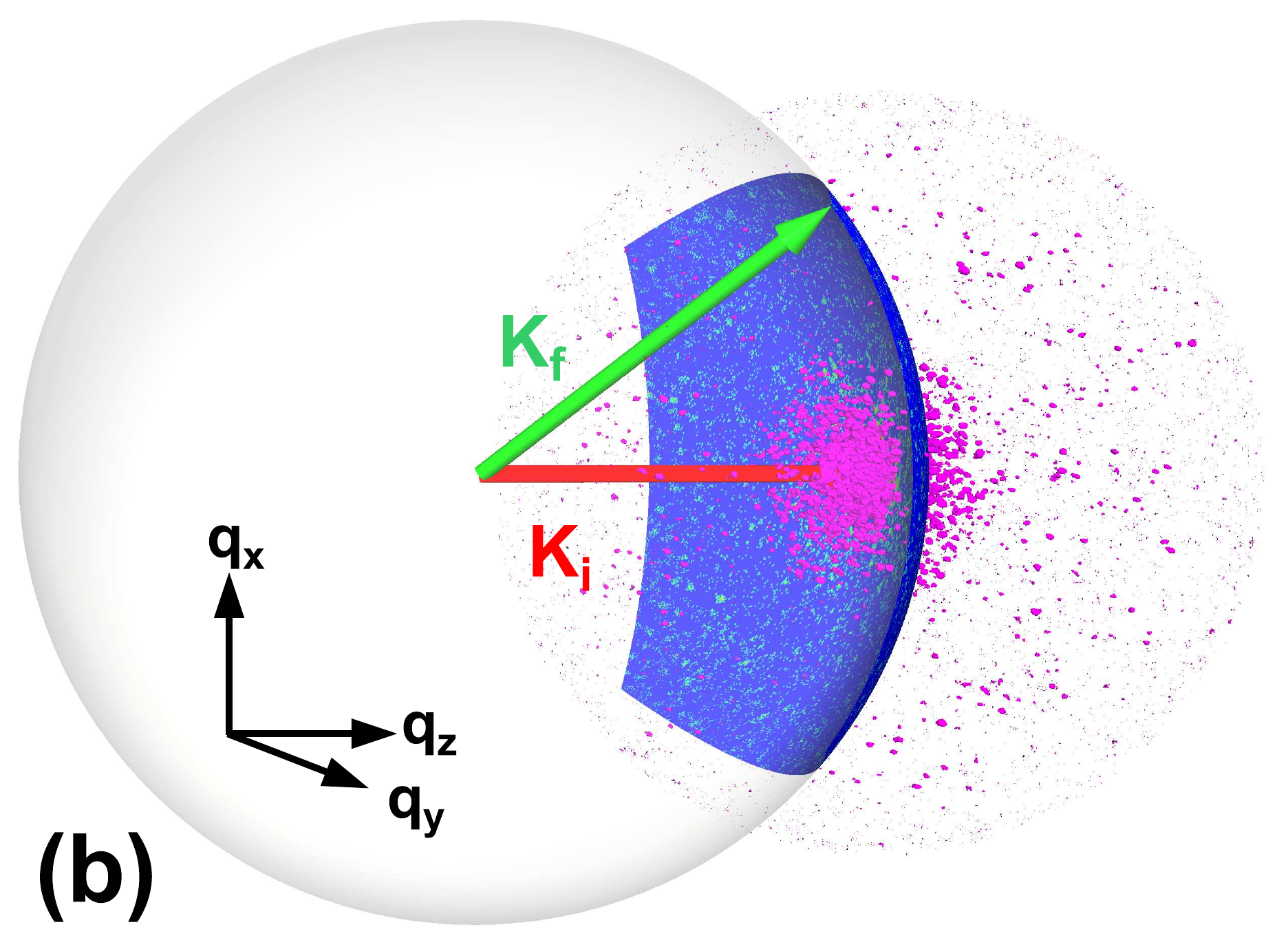}}
\caption{(Color online) Schematic view of the experimental geometry. (a) In real space, diffraction pattern from a sample in random orientation is measured by a single FEL pulse. (b) In reciprocal space the measured diffraction pattern correspond to a cut of the 3D intensity distribution by an Ewald sphere sector. Vectors $\mathbf{K}_i$ and $\mathbf{K}_f$ denote the incident and diffracted wavevectors.}
\label{Schem}
\end{figure*}

\begin{figure}[p]
\centerline{\includegraphics[width=0.45\textwidth]{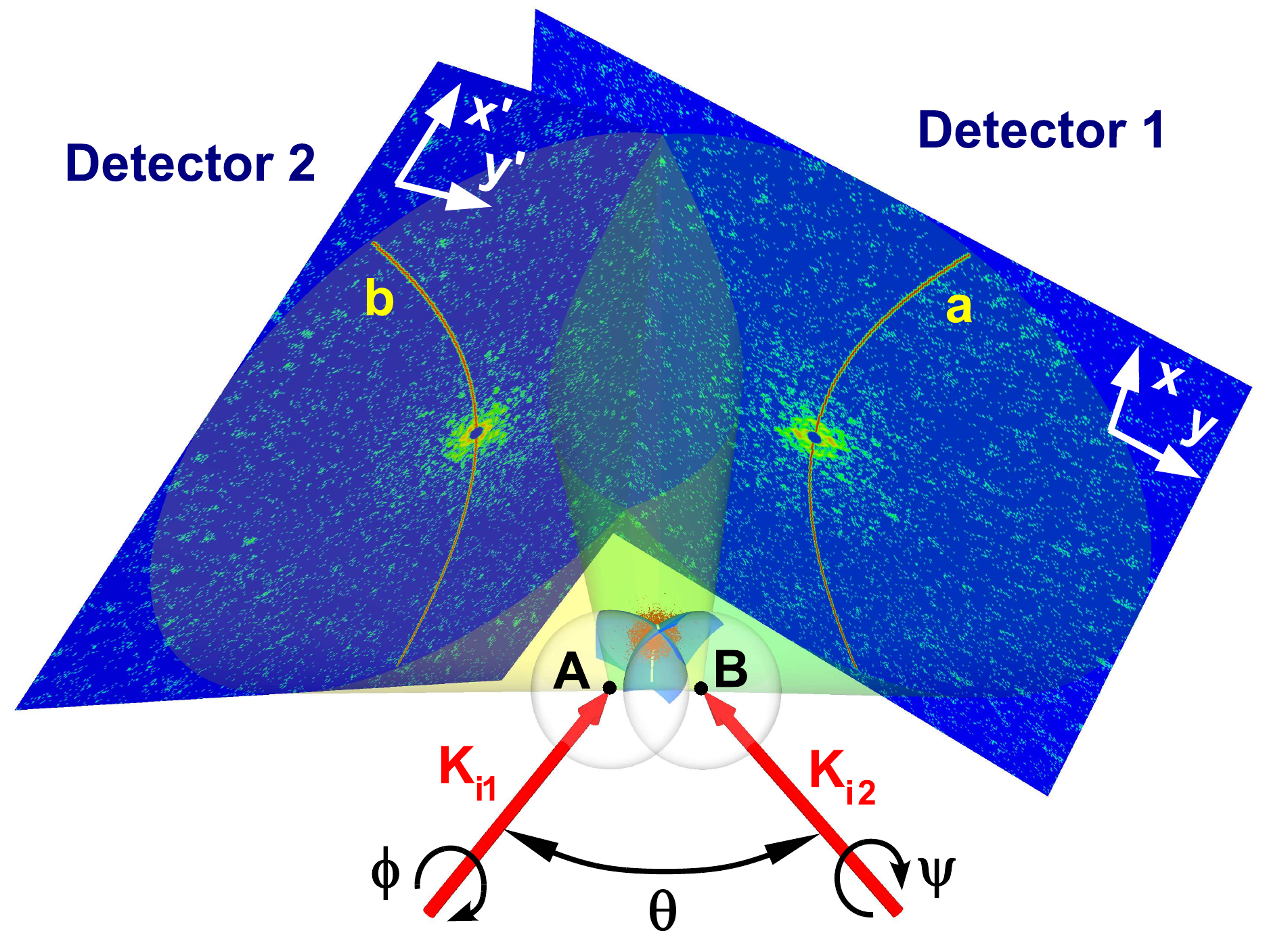}}
\caption{(Color online) Measurements of two reproducible samples at random orientation can be considered as two measurements of the same sample with two different incident beam directions indicated by vectors $\mathbf{K}_{i1}$ and $\mathbf{K}_{i2}$. Angles $\phi, \theta, \psi$ are Euler's rotation angles. Points A and B are the centers of the corresponding Ewald's spheres. Coordinates on the first and second detector are indicated as $x ,y$ and $x', y'$, respectively.}
\label{2Ewalds}
\end{figure}

In a typical single particle diffraction imaging experiment, a sample with unknown orientation is injected into the focused coherent x-ray beam of an FEL (Fig. \ref{Schem},a). The scattered radiation is measured in the far field by a 2D detector. This diffraction pattern can be mapped on an Ewald sphere \cite{AlsNielsen}, and represents a 2D cut of the 3D intensity distribution in reciprocal space (Fig. \ref{Schem},b). Alternatively, the diffraction pattern can be considered as a perspective projection of an Ewald sphere sector onto the 2D detector plane as viewed from the sample position (Fig. \ref{2Ewalds}).

Our previous studies suggest \cite{Mancuso2010} that, in order to increase the scattered signal, it is favorable to use longer x-ray wavelengths,
since the x-ray scattering cross-sections are larger at these wavelengths. At the same time the energy of the incident x-rays should be sufficient to penetrate the sample. To achieve high resolution the detector should also cover high scattering angles. Under these conditions a large sector of an Ewald sphere is covered, which is beneficial for orientation determination.

When two independent measurements of identical particles with different orientations are considered, the orientation of the first particle can be fixed as known. The orientation of the second particle can be uniquely described relative to the first one. Alternatively, two measured diffraction patterns could be considered to originate from the {same} particle in {different} experimental geometries. In this case the particle orientation is fixed, but the direction of the incident beam and the detector orientation are different for each measurement as shown in Fig. \ref{2Ewalds}. For the first measurement the incident beam direction, given by its wavevector $\mathbf{K}_{i1}$, can be taken along the $\mathbf{q}_z$ axis in the reciprocal space coordinate system shown in Fig. \ref{Schem},b. The direction of the incident beam for the second measurement is given by its wavevector $\mathbf{K}_{i2}$ (Fig. \ref{2Ewalds}). The relative orientation of the second geometry with respect to  the first one can be described  by three Euler angles $\phi, \theta, \psi$ \cite{LandauMex}. The choice of Euler angles is convenient, since rotations around the angles $\phi$ and $\psi$ in reciprocal space are equivalent to rotations by the same angles of detectors one and two in real space, respectively.

\begin{figure}[p]
\centerline{\includegraphics[width=0.4\textwidth]{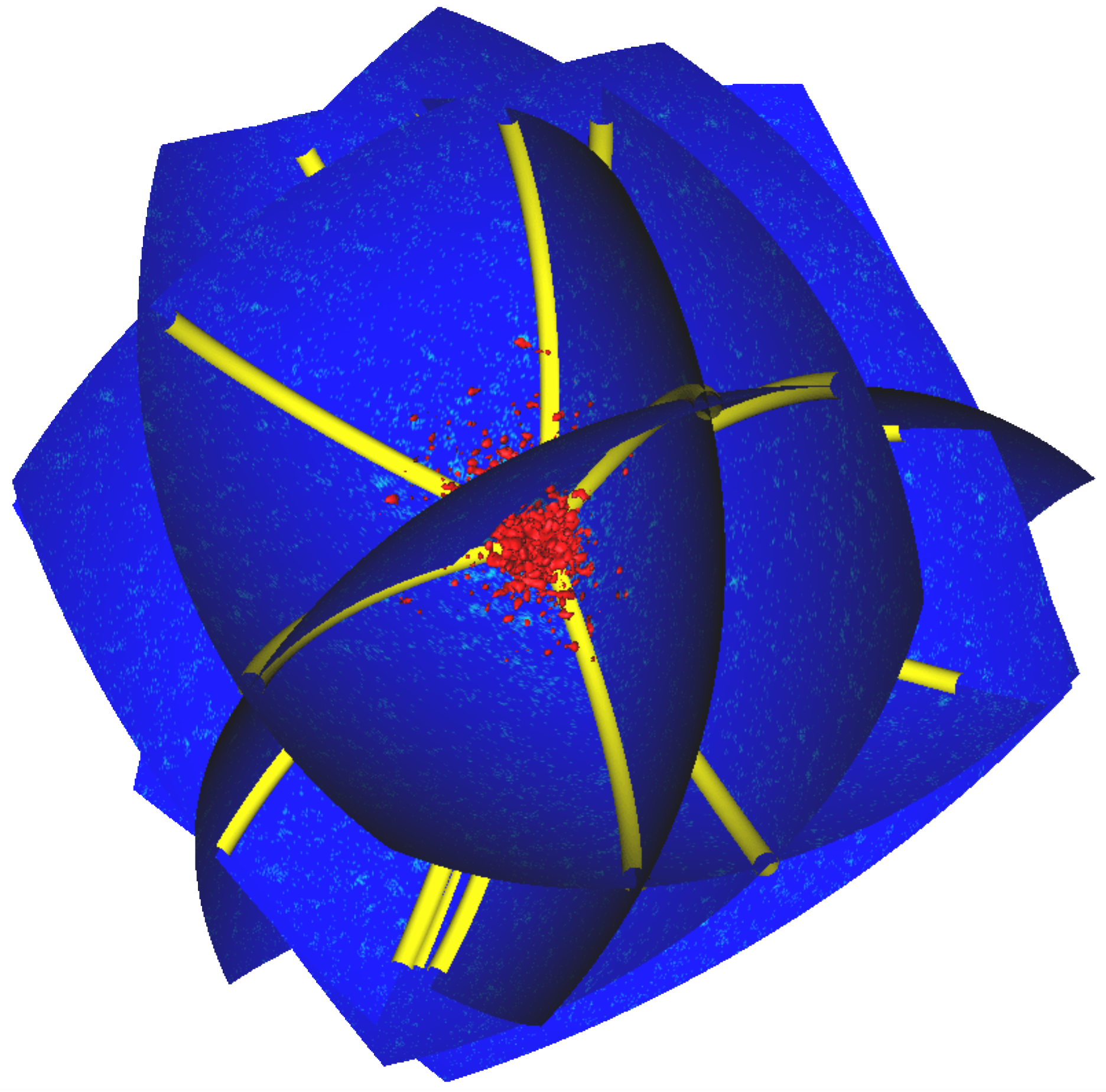}}
\caption{(Color online) Few Ewald sphere sectors intersecting the 3D intensity distribution of the sample in reciprocal space. Yellow lines indicate common arcs between different patterns.}
\label{ManyEwalds}
\end{figure}

For monochromatic x-rays the Ewald sphere has the radius $K = 2\pi / \lambda$, where $\lambda$ is the wavelength of the incident radiation. The Ewald spheres corresponding to the two measurements pass through the origin of the reciprocal space coordinate system (Fig. \ref{2Ewalds}).
The origin of the first Ewald sphere (see point A in Fig. \ref{2Ewalds}), for the incident vector $\mathbf{K}_{i1}$ is at $(0,0,-K)$  and the origin of the second sphere (point B in Fig. \ref{2Ewalds}), for the incident vector $\mathbf{K}_{i2}$, is at $(q_{x0},q_{y0},q_{z0})$ . The coordinates $q_{x0}, q_{y0}, q_{z0}$ are determined by a rotation of the point $(0,0,-K)$ around the reciprocal space origin $(0,0,0)$ by the Euler angles $\phi$ and $\theta$.
The intersection of the two spheres is a common arc that also passes through the origin of reciprocal space (see Fig. \ref{ManyEwalds},a). This common arc is projected on the two detectors (curves $a$ and $b$ in Fig. \ref{2Ewalds}). It is clear from this construction that the intensity along these arcs must be the same at both detectors. By analyzing the intensity correlations along all possible common arcs, the unique relative orientation of the two measurements can be determined.

It should be noted that a common arc can fix the relative orientation of two patterns only for experimental geometries with large scattering angle. Otherwise, the measured sector of the Ewald sphere can be considered as flat, and the common arc reduces to a straight line. This common line fixes only the angles $\phi$ and  $\psi$, but not the angle $\theta$, therefore a simultaneous analysis of at least three diffraction patterns is needed in this case \cite{VanHeel1987}.

The projection of the common arc on the first detector (curve $a$ in Fig. \ref{2Ewalds}) can be expressed in the detector 2D coordinate system $(x,y)$ by the following equation (see for details Appendix A)
\begin{eqnarray}\label{eq1}
&(&q_{x0}^2-(q_{z0}+K)^2) x^2 + (q_{y0}^2-(q_{z0}+K)^2) y^2 + \\   \nonumber
&&\; 2 q_{y0} q_{x0} x y + 2 d q_{x0} (q_{z0}+K) x + 2 d q_{y0} (q_{z0}+K) y = 0 \, ,
\end{eqnarray}
where $d$ is the sample-detector distance and  $x$, $y$ are the coordinates of the common arc projection on the first detector. Similar projection of the common arc on the second detector (curve $b$ in Fig. \ref{2Ewalds}) is also described by equation (\ref{eq1}) by substituting $x,y$ coordinates to $x'$, $y'=-y$.

As it follows from equation (\ref{eq1}), the curvature of the common arcs $a$ and $b$ at the detectors one and two is determined only by the angle $\theta$ and sample-detector distance $d$. Practically, the coordinates of the projections of common arcs at the detector planes are obtained by solving equation (\ref{eq1}) for each value of $\theta$ and $d$ with the fixed angles $\phi=\psi=0$. A set of curves corresponding to the fixed value of the angle $\theta$ and all other values of angles $\phi$ and $\psi$ is determined by rotation of the curve obtained on the previous step. This is implemented by rotation of the coordinate system $(x,y)$ corresponding to the first detector by angle $\phi$ and the coordinate system $(x',y')$ of the second detector by an angle $\psi$ (see Fig. \ref{2Ewalds} and Appendix B for details).

For each set of Euler angles and fixed sample-detector distance $d$ the coordinates of both arcs ($a$ and $b$ in Fig. \ref{2Ewalds}) are determined, and the intensities along these lines are compared by calculating the cross-correlation coefficient (CCC) $c^{ab}(\phi,\theta,\psi)$
\begin{equation}\label{eq3}
c^{ab}(\phi,\theta,\psi) = \frac{\sum_i{J^a_i(\theta,\phi) J^b_i(\theta,\psi)}}{\sqrt{\sum_i{[J^a_i(\theta,\phi)]^2}} \sqrt{\sum_i{[J^b_i(\theta,\psi)]^2}}},%
\end{equation}
\noindent where $J^{a,b}_i(\theta,\phi) = \ln[I^{a,b}_i(\theta,\phi)+1]$ are logarithms of the intensities $I^a_i(\theta,\phi)$ and $I^b_i(\theta,\phi)$ along the first ($a$) and second ($b$) arc.
The correct orientation of the second measurement with respect to the first one is given by the set of angles $\phi_B,\theta_B,\psi_B$ that maximize the CCC in Eq. (\ref{eq3}). To determine this orientation all three Euler angles ($\phi,\theta,\psi$) are varied sequentially with some angular step and CCCs for every orientation are calculated and compared.
This procedure is applied to all diffraction patterns until their orientation relative to the first one is determined (Fig. \ref{ManyEwalds},b).

It should be noted here that not only orientation but also the position of a particle in space should be taken into account explicitly in the analysis. The transverse position relative to the incoming x-ray beam adds only a phase to the scattered amplitude and scales its intensity. If each diffraction pattern is properly centered this does not cause any problems in the analysis since the phase is not recorded by the detector. The intensity of each diffraction pattern can be rescaled at the stage of composing the 3D intensity distribution in reciprocal space, as described later. At the same time, the particle-detector distance $d$, must be taken into account explicitly, due to its strong influence on the diffraction pattern. If two measurements are performed at different sample-detector distances $d_1$ and $d_2$, equation (\ref{eq1}) must be solved separately for both detectors taking into account the corresponding distances. This is especially important in real experimental conditions, when particles are injected in the beam, due to variations of the distance $d$ from shot to shot. We also assume in our analysis that the particle size is much smaller than any variations of beam intensity. More details on our practical implementation of the common arc algorithm are presented in Appendix B.

The common arc algorithm described in this section performs well for data sets with high signal to noise ratio \cite{Huldt2003}. However, it often fails in practical applications for a low number of scattered photons. One way to overcome this problem is presented in the following section.

\section{\label{sec:advanced}Advanced algorithm for orientation determination in the presence of noise}

In the previous section, the common arcs between one diffraction pattern (that we define as a base pattern) and all other diffraction patterns were analyzed. At the same time we should note that each diffraction pattern has a common arc with other diffraction patterns (see Fig. \ref{ManyEwalds}). Therefore, common arcs between all patterns could, in principle, be analyzed simultaneously. Such analysis can significantly improve the fidelity of the orientation, however, its practical implementation requires an increase in computational resources. A compromise can be found by implementing the following strategy. As a first step a set of base diffraction patterns $N_{base}$ is analyzed with respect to each other to determine the correct orientations of these chosen patterns. In the next step all other patterns are oriented with respect to each of these base patterns. This implementation requires $N_{base}$ times more calculations compared to a single base pattern.  In the final step all intensities are mapped to a 3D array of voxels in reciprocal space by 3D gridding and averaging procedure. The benefit of this approach is the possibility to solve the orientation problem for noisy data, as will be demonstrated in the following section (see also for a detailed discussion Appendix C).

In a real experimental situation all diffraction patterns have different intensities due to shot to shot intensity jitter of the FEL and the fact that each injected particle is hit by a different part of a focused beam. As a consequence all measured diffraction patterns have to be rescaled. This is implemented in the algorithm by utilizing the fact that each of two patterns have a common arc and that the intensities along this arc must be equal. The scaling factor for the intensities can be determined by taking the ratio of intensities of two diffraction patterns along the common arc. Having all information about the experimental geometry, orientation, and scaling factor for each pattern the 3D intensity distribution in reciprocal space can be constructed.

The orientation determination can be significantly improved by an additional refinement that is based on the correlations between an individual pattern and the whole 3D intensity distribution. It can be implemented in the following way.
First, the 3D intensity distribution is obtained from all but one selected diffraction pattern. Then the orientation of the selected pattern is varied in a small angular range and the correlation between this 2D pattern and the whole 3D intensity distribution is analyzed. The orientation corresponding to the highest correlation value is considered to be the correct one. Then, the rescaled intensity of the selected pattern with the refined orientation is included in the new 3D intensity distribution in which the next diffraction pattern is excluded and the refining procedure is repeated. By applying this approach to all diffraction patterns the final 3D intensity distribution is obtained. This procedure can also be applied to identify diffraction patterns from "wrong" particles (the ones that do not belong to a set of samples under investigation). Clearly, correlation coefficients of diffraction patterns originating from these particles and the whole 3D intensity distribution will be quite low, which can be used as a criteria for rejection of these diffraction patterns from the future analysis.

If the structure has a known symmetry, this can be used as an additional constraint for orientation determination \cite{Vainstein1986,VanHeel1987}. Using symmetry conditions each diffraction pattern can be oriented individually with respect to the selected symmetry axis. This is contrary to the structures without symmetry when at least two patterns are required for orientation determination. Applying symmetry conditions it is possible to get sufficient number of diffraction patterns for 3D representation of the scattered intensity in reciprocal space even with a limited data set or a large area of missing data due to a big beamstop. This approach was successfully used for simulated data for a sample with icosahedral symmetry discussed in the next section as well as for experimentally measured diffraction patterns of a Mimi virus obtained in coherent diffraction imaging experiment at FLASH \cite{Mimi}.

It is interesting to note that presented implementation of the common arc algorithm also allows to determine the unknown symmetry of the object. This can be obtained by the analysis of angular orientations appearing with the highest probability. Such orientations can be found in a 3D angular map ($\phi$, $\theta$, $\psi$) of all possible orientations and reveal themselves as regions with high density (see for details Appendix C). For example, for structures with icosahedral symmetry it will correspond to 120 most likely orientations in reciprocal space that are related to the icosahedral symmetry transformation matrix.

A sampling rate of at least two in each direction in diffraction pattern is required for a successful implementation of the algorithm described here. The same requirement is valid for the phase retrieval algorithms applied for the reconstruction of electron density of the samples. A higher sampling rate is beneficial for orientation determination because each speckle consists of more pixels. At the same time data with a lower sampling rate have a higher signal in each pixel, which could become important for orientation determination of data with a low signal to noise ratio \cite{Mancuso2009}. By testing different sampling conditions we found that an optimal sampling rate is in the region from two to three. In practice, to increase signal the experimental data could be binned for orientation determination, while the reconstruction is performed on the original unbinned data set. Applying on a final step phase retrieval algorithms \cite{Fienup1982,Marchesini2007} to the 3D data set of the intensity distribution, the electron density of the sample can be obtained.

\section{\label{sec:test}Numerical test of the algorithm}

\begin{figure}[tbp]
\centerline{\includegraphics[width=0.45\textwidth]{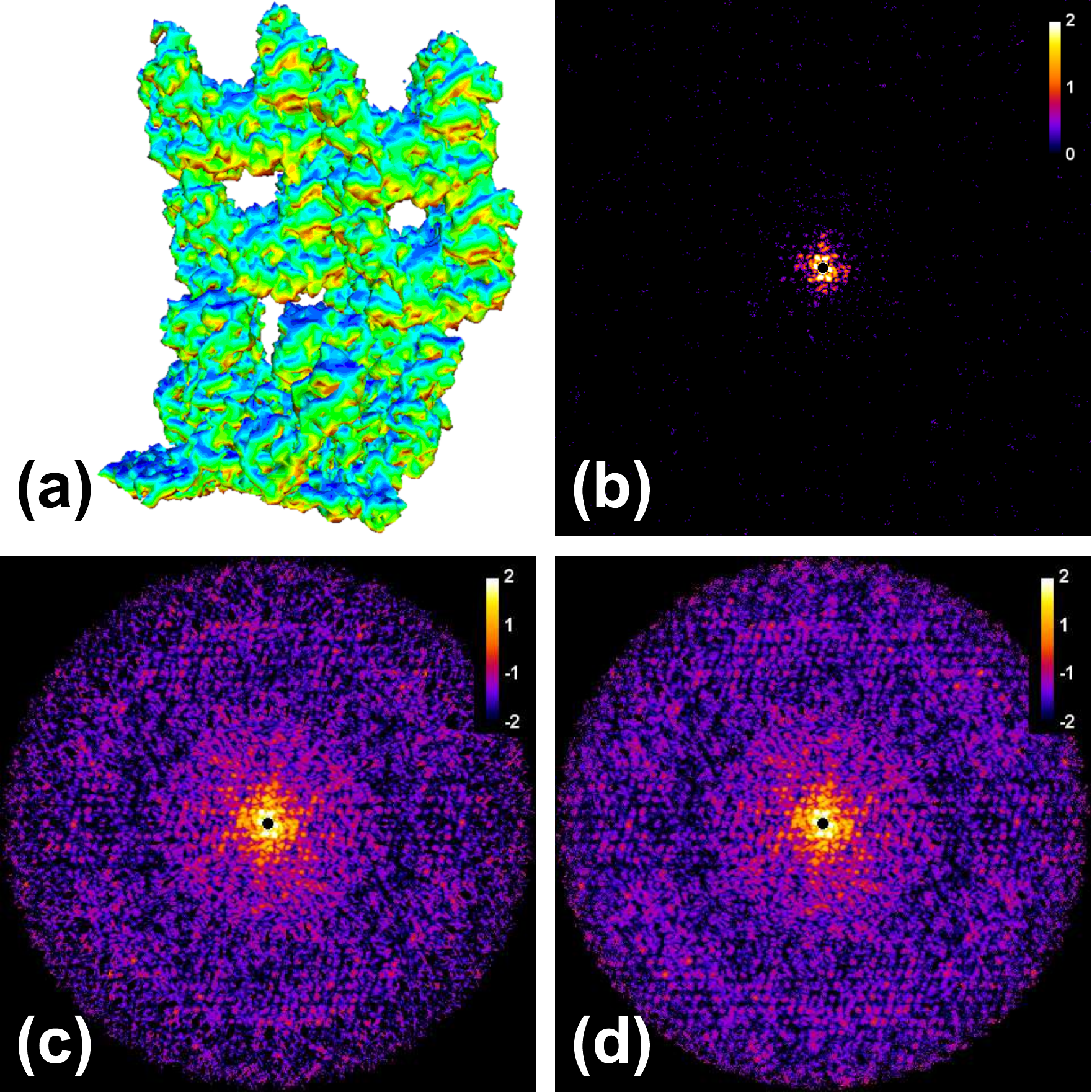}}%
\caption{(Color online) Artificial protein structure without any symmetry combined from the 2BTV and 8RUC macromolecular structures. (a) Iso-surface of the electron density, (b) single diffraction pattern (edge resolution 3.92{\AA}), (c,d) 2D central cuts (edge resolution 3.3{\AA}) through the constructed 3D intensity distribution in reciprocal space for the patterns with a known orientation (c), and the patterns with the orientations determined using the proposed algorithm (d). All diffraction patterns are presented in logarithmic scale.}
\label{Ruc_result}
\end{figure}

The algorithm was tested with two different biological structures. The first one was an artificial protein structure without any symmetry combined from the 2BTV and 8RUC macromolecular structures \cite{PDB} (see Fig. \ref{Ruc_result},a). It has a size of $13\times 19\times 28$ nm$^3$ and consists of about 124 000 non-hydrogen atoms. The second one was a human adenovirus penton base 2 12 chimera 2c6s \cite{PDB}. It has icosahedral symmetry with the diameter of 27 nm and consists of about 200 000 non-hydrogen atoms (Fig. \ref{2c6s_result},a).

Diffraction patterns for both structures where calculated \footnote{All diffraction patterns were simulated using the program \texttt{moltrans}.} at 3 {\AA} wavelength, with a detector size of 100 mm and a sample-detector distance of 50 mm, providing the maximum scattering angle of $45^\circ$. The achievable resolution in this geometry was 3.92 {\AA} at the detector edge and 3.3 {\AA} at its corner. The number of detector pixels was $512\times 512$ for the first sample (providing minimum sampling rate of 2.5) and $360\times 360$ for the second (with a sampling rate of two). The incoming flux was $10^{12}$ photons focused uniformly on a $100\times 100$ nm$^2$, and Poison noise was added to each diffraction pattern. The average flux at the edge of the detector was 0.05 and 0.15 photons per pixel corresponding to 0.45 and 0.6 photons per Shannon angle for the first and second structure, respectively. For the first structure $36\times36\times18=$23 328 patterns were simulated with a 10$^\circ$ increment for each Euler angle. For the second structure 12 000 randomly oriented patterns were simulated. A beamstop with diameter of about 2 mm covering 1.5 speckles was introduced in all simulated diffraction patterns, and this region was excluded from the calculation of correlation coefficients. Typical diffraction patterns for a single FEL pulse simulated in the experimental conditions described above are shown in Fig. \ref{Ruc_result},b and Fig. \ref{2c6s_result},b for the first and second structure, respectively.

\begin{figure}[tbp]
\centerline{\includegraphics[width=0.5\textwidth]{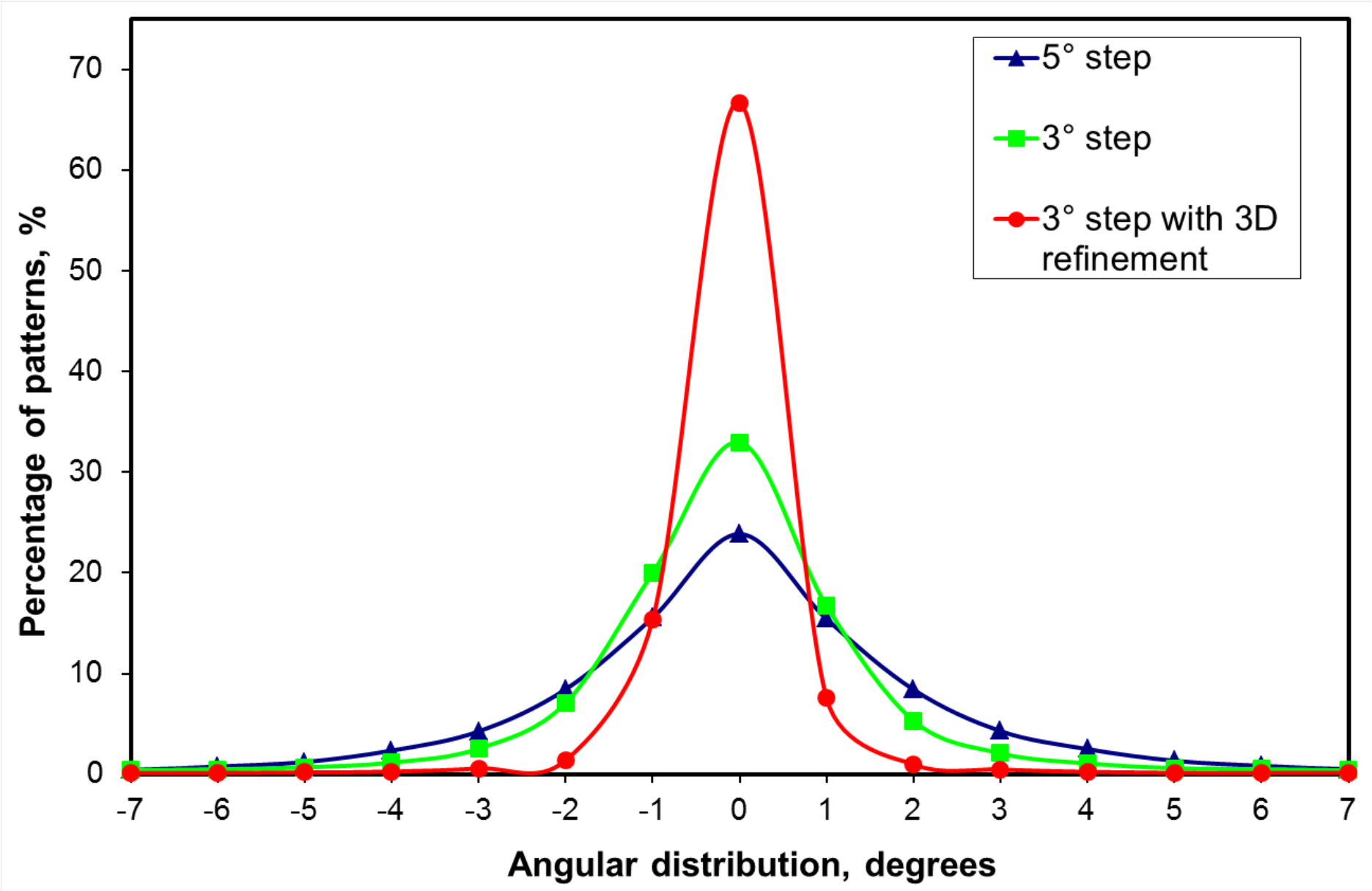}}%
\caption{(Color online) Distribution of the angular error of the determined orientations for the structure without symmetry. Blue line corresponds to 5$^\circ$ angular step, green line 3$^\circ$, and red line 3$^\circ$ after the 3D refinement (see text for details).}
\label{RightAnglesPlot}
\end{figure}

The correct orientation of each diffraction pattern was determined using the algorithm described in the previous sections. This allowed us to obtain the full 3D intensity distribution in reciprocal space for each sample. A central slice through this distribution constructed from the oriented diffraction patterns corresponding to the first structure is presented in Figure \ref{Ruc_result},d. For comparison, the same slice through the 3D intensity distribution obtained from the known orientation of each diffraction pattern is shown in Fig. \ref{Ruc_result},c. It is well seen that the slice obtained as a result of orientation determination well reproduces all features of an "ideal" intensity distribution, small deviations can be attributed to angular misalignment. This misalignment between the angles obtained from the common arc algorithm and the correct angles for the first structure is presented as a plot in Fig. \ref{RightAnglesPlot}. The accuracy of the orientation determination correlates strongly with the angular step size for the Euler angles ($\phi$, $\theta$, $\psi$). Clearly, a finer angular step size requires more computational time that scales as a third power of the step size. It is well seen in Fig. \ref{RightAnglesPlot} that a three degree angular step being five times slower still gives higher accuracy in orientation determination comparing to a five degree step. An additional improvement in the angular determination is obtained by the final orientational refinement of each diffraction pattern with respect to 3D intensity distribution, as described in the previous section (see Fig. \ref{RightAnglesPlot}).

It is interesting to observe how the signal is increased by the number of diffraction patterns used in the analysis. In Fig. \ref{2c6s_result},d a central slice through the 3D array
representing the number of patterns contributing to each voxel of the constructed 3D intensity distribution is presented. It can be seen from this figure that at least 100 patterns from the analyzed 12 000 contribute to each voxel inside a resolution ring of 4 {\AA}. One more intriguing feature can be observed in this figure.  Though the initial diffraction patterns were simulated up to 3.92 {\AA} resolution at the edge of the detector the 3D intensity distribution obtained from the algorithm has distinguishable features up to 3.3 {\AA} resolution (see dark outer ring in Fig. \ref{2c6s_result},d). This additional signal comes from the corners of the diffraction patterns.

\section{\label{sec:level1}Summary}

In summary, we proposed a method for the angular orientation determination in single particle coherent imaging experiments based on the common arc algorithm. We obtained a significant improvement of this approach by introducing a simultaneous analysis of the common arcs for several diffraction patterns. This gives the possibility to apply the method to data with a low level of signal to noise ratio as well as to skip classification step which can reduce achievable resolution. Additionally, we proposed an orientational refinement of diffraction patterns that can improve the quality of the final 3D intensity distribution in reciprocal space.

The algorithm proposed here has several advantages compared to other approaches \cite{Fung2009, Elser2009}. It scales linearly with the number of measured patterns and total number of pixels in the diffraction patterns. The algorithm is easy to parallelize, because most of the cross-correlation analysis is performed between pairs of independent diffraction patterns. It has minimum memory requirements, because the data can be processed in parts.

We foresee that this approach has the potential to be the key for the success in the analysis of single particle diffraction imaging experiments and will allow to reach sub-nanometer resolution in three-dimensional imaging of biological specimens.

\begin{acknowledgments}
We are thankful to E.~Weckert for a permanent interest and support of this project, as well as for the use of program \texttt{moltrans} for simulation of diffraction patterns from biological structures, and to A. Singer and U. Lorenz for a careful reading of the manuscript. Part of this work was supported by BMBF Proposal 05K10CHG `'Coherent Diffraction Imaging and Scattering of Ultrashort Coherent Pulses with Matter`' in the framework of the German-Russian collaboration `'Development and Use of Accelerator-Based Photon Sources`' and
the Virtual Institute VH-VI-403 of the Helmholtz association.
\end{acknowledgments}

\appendix




\section{Derivation of the main equations}

\begin{figure}[tbp]
\centerline{\includegraphics[width=0.45\textwidth]{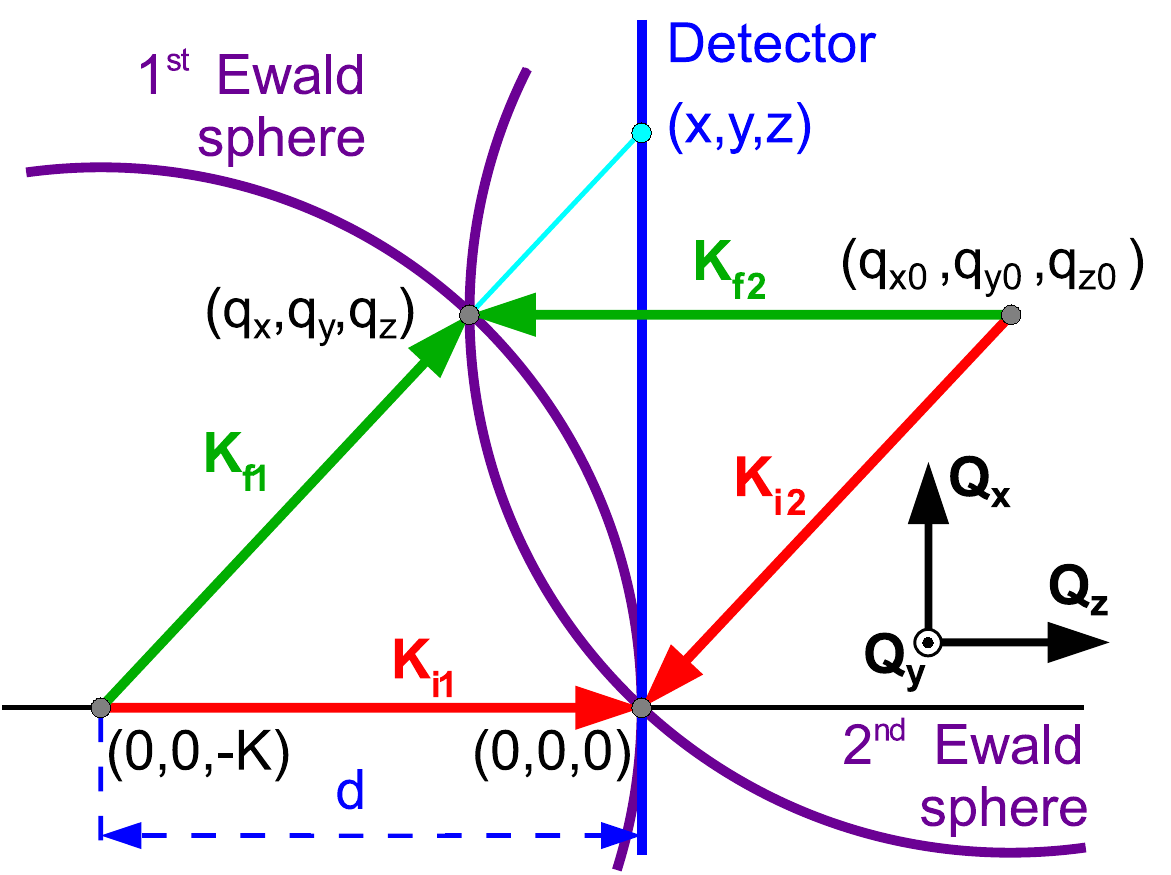}}
\caption{(Color online) Schematic view of the intersection of two Ewald spheres.}
\label{SchemView}
\end{figure}

Equation (\ref{eq1}) was derived using the following considerations.
Both intersecting Ewald spheres (Fig. \ref{SchemView}) have radii equal to the wave vector $K = 2 \pi / \lambda$ ($|\mathbf{K_i}|^2=|\mathbf{K_f}|^2=K^2$), where $\lambda$ is wavelength of the incident radiation. Therefore the coordinates ($q_x, q_y, q_z$) of the intersection curve must satisfy the equations
\begin{eqnarray}
  &q_x^2 + q_y^2 + (q_z+K)^2 = K^2,& \label{eqA2}\\
  &(q_x - q_{x0})^2 + (q_y - q_{y0})^2 + (q_z - q_{z0})^2 = K^2.& \label{eqA3}
\end{eqnarray}

 The center of the second Ewald sphere $(q_{x0}, q_{y0}, q_{z0})$ lies on the distance $K$ from the center of reciprocal space ($|\mathbf{K}_{i2}|^2=K^2$), therefore
\begin{equation}\label{eqA4}
  q_{x0}^2 + q_{y0}^2 + q_{z0}^2 = K^2.
\end{equation}

From equations (\ref{eqA2} - \ref{eqA4}) the formula describing intersection of two Ewald spheres can be derived:
\begin{equation}\label{eqA5}
    (q_{y0}^2 + q_{x0}^2) q_y^2 + 2 q_{y0} q_{z0} (q_z - K) q_y + (q_{z0}^2 + q_{x0}^2) q_z^2 +K^2 (q_{z0}^2 - q_{x0}^2) - 2 K q_{z0}^2 q_z = 0.
\end{equation}

As soon as the diffracted vector $\mathbf{K}_{f1}$ (Fig. \ref{SchemView})  has the same direction in both real and reciprocal spaces (angles coincide) the following relation between coordinates of a pixel on the detector $(x,y,z)$ in real space and corresponding coordinates of the end of $\mathbf{K}_f$  $(q_x,q_y,q_z)$ in reciprocal space can be written:
\begin{equation}\label{eqA1}
    \frac{x}{q_x} = \frac{y}{q_y} = \frac{z}{q_z}.
\end{equation}

As soon as the distance from the sample to the detector ($d$) is fixed, $z \equiv d$. Equation (\ref{eq1}) can be easily derived from equations (\ref{eqA5}, \ref{eqA1}).



\section{\label{sec:common1}Common arc algorithm}

Due to the properties of Euler angles, angle $\phi$ ($0\leq \phi < 2\pi$) can be attributed to the rotation of the reciprocal space coordinate system around the incident beam ($\mathbf{K}_{i1}$), angle $\theta$ ($0\leq \theta < \pi$) corresponds to the rotation around the new position of vector $\mathbf{q}_y$ and angle $\psi$ ($0\leq \psi < 2\pi$) is the final rotation of the coordinate system around the new position of the vector $\mathbf{q}_z$ - vector $\mathbf{K}_{i2}$ in Fig. \ref{2Ewalds}.

In practice, the curvature of the common arcs $a$ and $b$ in Fig. \ref{2Ewalds} is determined only by the angle $\theta$ and distance $d$. The coordinates of the projections of common arcs on detector planes can be obtained for each value of $\theta$ and $d$, with the fixed angles $\phi = \psi = 0$, by solving equation (\ref{eq1}). Other curves (for all values of angles $\phi$ and $\psi$) at the fixed value of angle $\theta$ are determined by rotation of the curve obtained at the previous step. Coordinate system $(x,y)$, corresponding to the first detector, is rotated by an angle $\phi$ and the coordinate system $(x', y')$, corresponding to the second detector, by an angle $\psi$ (Fig. \ref{2Ewalds}).

\begin{figure}[tbp]
\centerline{\includegraphics[height=0.5\textheight]{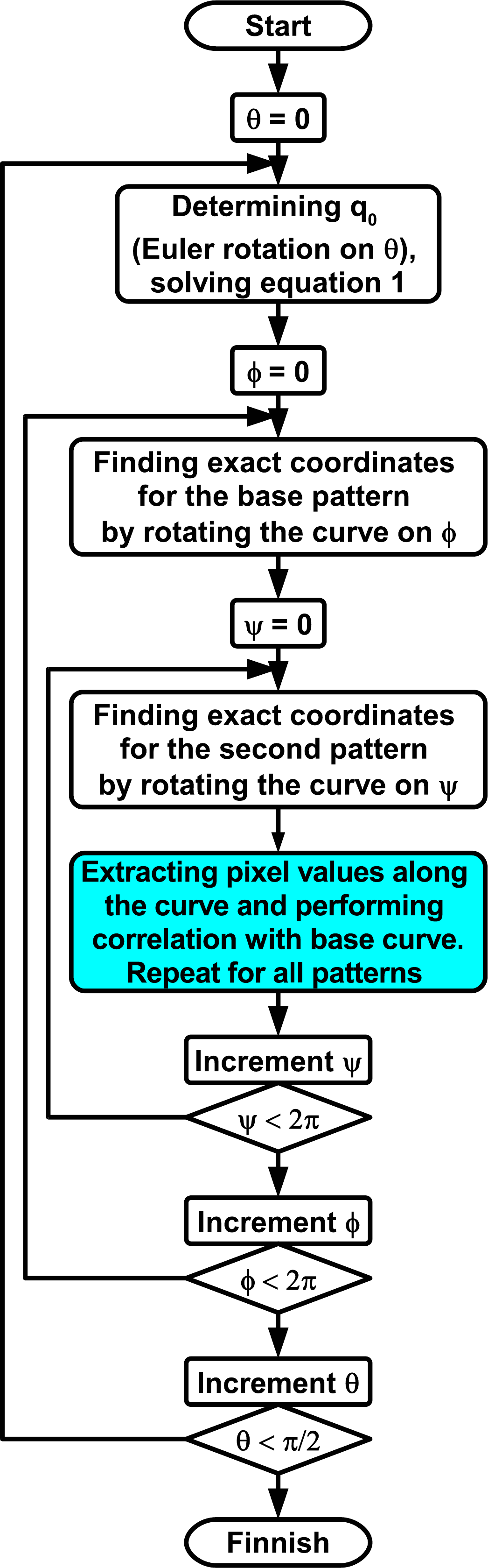}}
\caption{Flowchart for efficient data analysis with common arc algorithm.}
\label{Algorithm}
\end{figure}

The common arc algorithm described in this paper was implemented using the following scheme (supplementary Fig. \ref{Algorithm}). Calculation starts with fixed angles $\phi=0$ and $\psi=0$. Then coordinates $q_{x0}$, $q_{y0}$, $q_{z0}$ are calculated by Euler rotation on $\theta$ angle of the vector $\mathbf{K}_{i1}$ (point $(0,0,-K)$). After this equation (\ref{eq1}) is solved and coordinates (x,y) for the common arc are found. This curve is rotated on angle $\phi$ for the first pattern (each pair of $(x,y)$ is multiplied by corresponding rotation matrix) and on the angle $\psi$ for the second pattern (with exchange $y \rightarrow -y$). After full determination of the curves for both patterns corresponding values of intensities (along the curve) are extracted using the interpolation described below. The intensities along the curves are then correlated. This process continues for all angles $\psi$ in the region $0 \leq \psi < 2 \pi$ and for all angles $\phi$ in the region $0 \leq \phi < 2 \pi$. Then the whole process is repeated for different $\theta$ value.

The common arcs approach has difficulties when angle $\theta$ is large. In this case the intersection between two spheres reduces to a closed circle. When angle $\theta$ approaches $\pi$ this circle shrinks to a point at the origin of reciprocal space coordinates $(0,0,0)$. Therefore at large $\theta$ (close to $\pi$) the projection of the common arc to the detector plane, described by the equation (\ref{eq1}), degenerates to an ellipse. Therefore the length along an arc and a circle can be different. Moreover the correlation coefficients found for the arcs with different curvature can hardly be compared, because the ends of such curves correspond to different $q$-range and so some curves will have good signal at the ends and some - mostly noise. From this considerations it is clear that it is difficult to compare curves obtained for small and big $\theta$ angles. To solve this problem we limited the range of acceptable $\theta$ angles to the range $0 \leq \theta \leq \pi/2$. To cover the range of angles $\pi/2 < \theta < \pi$ we used the fact that reciprocal space is centro-symmetric for scattering on non-absorbing objects (Friedel's law).
To take this into account for angles $\pi/2 < \theta < \pi$ we invert direction of the vector $\mathbf{K}_{i2}$ (Fig. \ref{2Ewalds}) to its opposite $-\mathbf{K}_{i2}$, that corresponds to the following transformation: rotation of $\phi$ by $\pi$ ($\phi \rightarrow \phi + \pi$), rotation of $\theta$ by $\theta \rightarrow \pi - \theta$ and final rotation of $\psi$ by $\pi$ ($\psi \rightarrow \psi + \pi$). To finish inversion transformation we change $y \rightarrow -y$ (equation \ref{eq1}) in detector plane.


To find the intensities corresponding to each point of the curve described by equation (\ref{eq1}), some sort of gridding must be performed. In our calculations we used interpolation in the form of an average of four nearest neighbor pixels. We checked also other schemes of interpolation: nearest neighbor and bilinear interpolation \cite{Wolfram}. The first one is faster but lacks accuracy, the second is computationally much slower without noticeable increase in quality.

All common arcs for different $\theta$ values (different curvature) should cover the same q-distance in reciprocal space. Therefore the step between points on the curve should remain constant. For this reason equation (\ref{eq1}) was differentiated analytically and starting from the center of detector ($x=y=0$) each next point on the curve is calculated keeping the distance $(dx^2+dy^2)=const$.

The accuracy of calculations of cross-correlation coefficient (equation \ref{eq3}) for all patterns can be increased by replacing intensities $I(q)$ with their logarithms, more precisely by $\ln(1+I(\mathbf{q}))$. Also for noisy data the following form of CCC is more beneficial:

\begin{equation}\label{eqcc3}
c^{ab}(\phi,\theta,\psi) = \frac{\sum{(J^a(\mathbf{q_i}) -<J^a(|\mathbf{q_i}|)>) (J^b(\mathbf{q_i}) -<J^b(|\mathbf{q_i}|)>)}} {\sqrt{\sum{(J^a(\mathbf{q_i}) -<J^a(|\mathbf{q_i}|)>)^2}} \sqrt{\sum{(J^b(\mathbf{q_i}) -<J^b(|\mathbf{q_i}|)>)^2}}},%
\end{equation}

where $J(\mathbf{q}) = \ln(1+I(\mathbf{q}))$ and $<J(|\mathbf{q_i}|)>$ is a radial averaged value of intensity corresponding to a ring with the radius $|\mathbf{q_i}|$ on a diffraction pattern.




\section{\label{sec:advanced1}Advanced algorithm for processing noisy data}

\begin{figure}[tbp]
\centerline{\includegraphics[width=0.45\textwidth]{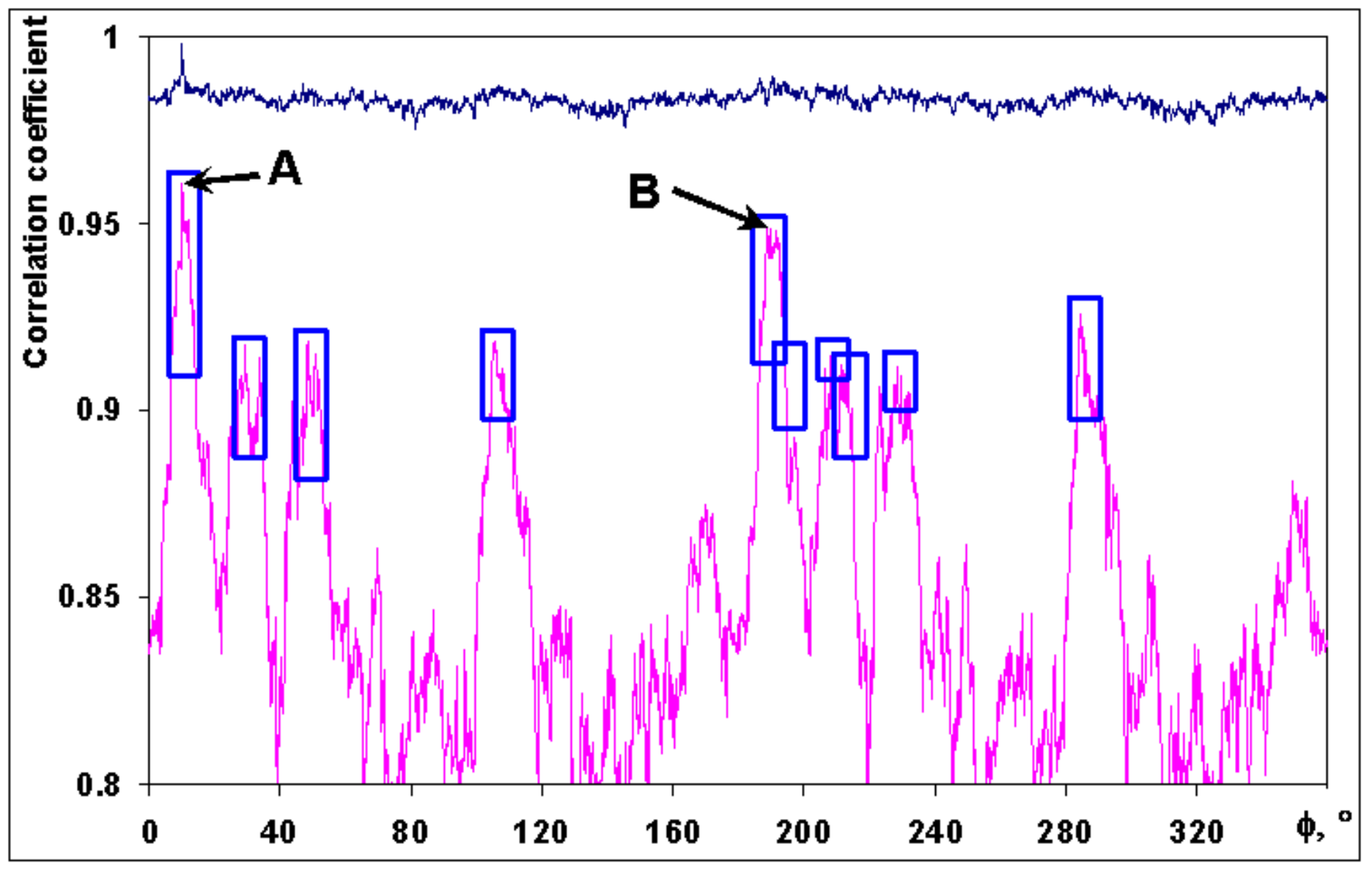}}
\caption{(Color online) Correlation coefficient between two patterns for different $\phi$ angles. Upper curve for ideal data, the lower one for the noisy data.}
\label{PlotCor}
\end{figure}

Fig. \ref{PlotCor} shows typical correlations, calculated with equation (\ref{eq3}), between two noisy diffraction patterns for different angles $\phi$ with fixed angles $\theta$ and $\psi$ (bottom curve). For comparison, correlation coefficients corresponding to a perfect data set are plotted in the same Fig. \ref{PlotCor} (top curve). The best correlation coefficient between two noisy patterns can correspond to completely wrong orientation (like point B in Fig. \ref{PlotCor}). Therefore several orientations ($N_{angl}$) corresponding to the best set of correlations must be stored. To avoid storage of almost identical angles, some tolerance $A_{tol}$ in the best orientation angles determination is necessary. Only one angle in the range $\pm A_{tol}$ with the highest correlation coefficient is stored. This leads to storage of only one angle per tolerance region (rectangles in Fig. \ref{PlotCor}). Therefore only one point per marked rectangle in Fig. \ref{PlotCor} is stored in the list of "best" angles (for example in Fig. \ref{PlotCor} $N_{angl}=10$ and $A_{tol}=5^\circ$).

The algorithm of orientation determination is following. In the first step all base patterns are correlated to each other using the algorithm described in the section \ref{sec:common}. As a result, $N_{angl}$ "best" angles between each pattern and all other $N_{base}-1$ base patterns are stored. Therefore each pattern has $(N_{base}-1) N_{angl}$ "best" angles with respect to all other base patterns. Then all these angles are recalculated with respect to one selected pattern (which attributed all angles equal to 0) - we shall call it the zeroth pattern. This is done in the following way: each pattern's base angles with respect to other bases are recalculated to angles with respect to zeroth pattern taking into account that each base has $N_{angl}$ angles with respect to zeroth. In this way $(N_{base}-2) N_{angl}^2 + N_{angl}$ angles for each pattern with respect to zeroth are determined.

\begin{figure}[tbp]
\centerline{\includegraphics[width=0.45\textwidth]{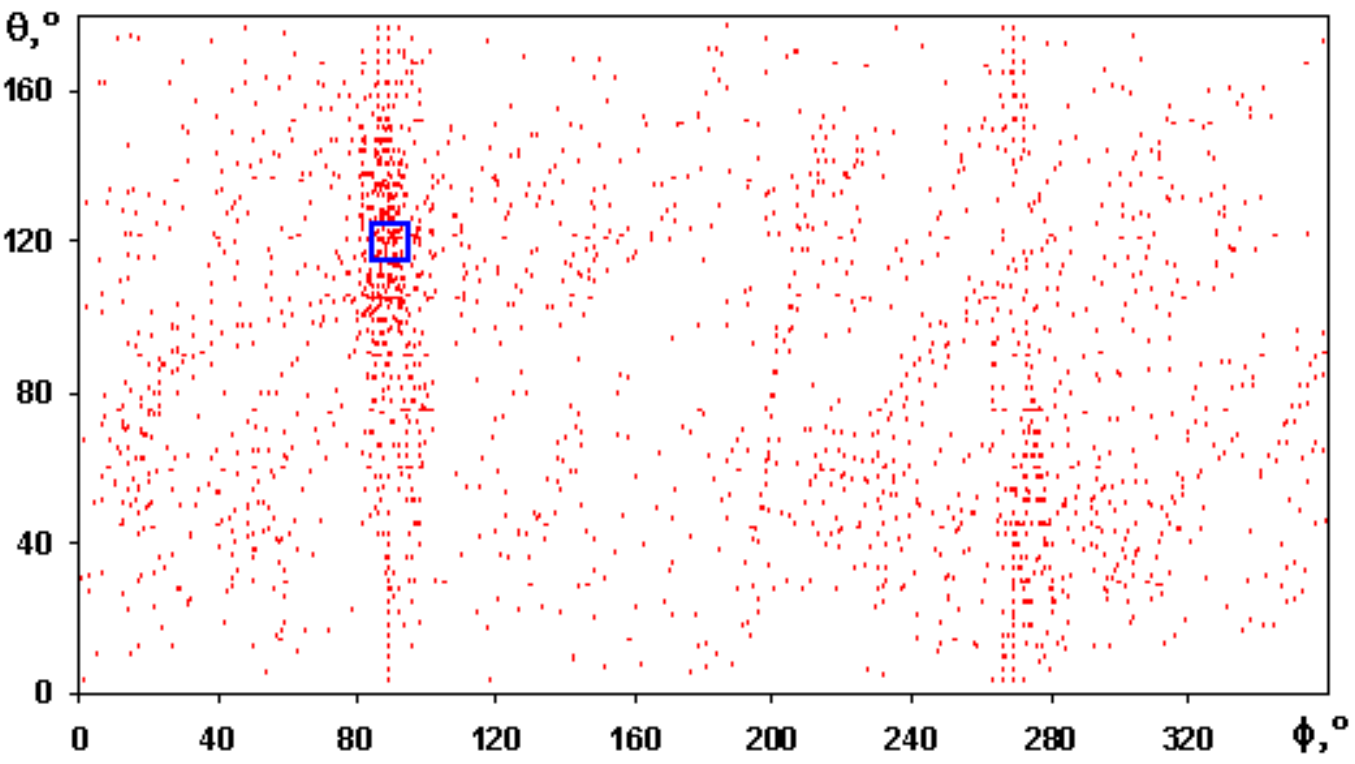}}
\caption{(Color online) Two-dimensional $(\phi, \theta)$ distribution of angles corresponding to good correlation between a pattern and all base patterns. A blue box correspond to the region with highest density of good orientations.}
\label{Plot2Dangles}
\end{figure}

Let's explain this step on an example. Consider one of the base patterns $P_i$. This pattern has $N_{angl}$ angles with respect to the zeroth pattern $P_0$. The pattern $P_i$ has also $N_{angl}$ angles with respect to the first pattern $P_1$. But the first pattern $P_1$ itself has $N_{angl}$ angles with respect to the $P_0$. So, pattern $P_i$ has already $N_{angl}+N_{angl}^2$ angles to the $P_0$. Then it also has $N_{angl}^2$ angles to the $P_0$ through the second pattern $P_2$. This process is continued for all base patterns. Finally all $(N_{base}-2) N_{angl}^2 + N_{angl}$ angles determined for one base pattern ($P_i$) are plotted in 3D space of angles $(\phi, \theta, \psi)$ and the angular region with the maximum density of points is selected as the best angle. In practice it is done in the following way: a number of points (in the tolerance region $\pm A_{tol}$ for each Euler's angle) near each point is calculated. The point with the biggest number of neighbors is considered to be the best estimate. Then the correct angle is determined by averaging of all positions of the neighbors. In Fig. \ref{Plot2Dangles} an example of such operation in 2D case (for angles $\theta$ and $\phi$ with fixed $\psi$) is presented. The best angle corresponds to the middle of the rectangle in Fig. \ref{Plot2Dangles}.

As soon as all correct angles for the base patterns are determined, all other patterns can be oriented with respect to known bases. At this step only $N_{base} N_{angl}$ angles need to be considered for each pattern under analysis in the algorithm described above.

For better orientation determination the base patterns should be selected carefully among all measured. All base patterns should have different orientations, more precisely different $\theta$ angles. Because initially all angles are unknown this requirement can be satisfied by the analysis of 2D correlations between different patterns. This is performed by calculating 2D correlation between pairs of patterns for different angles $\phi$ and the best correlation coefficient is stored. Then patterns with the worst 2D cross-correlations between each other are selected as bases. Also for experimental data analysis, base patterns should be selected according to the recorded signal quality.

\begin{figure}[tbp]
\centerline{\includegraphics[width=0.45\textwidth]{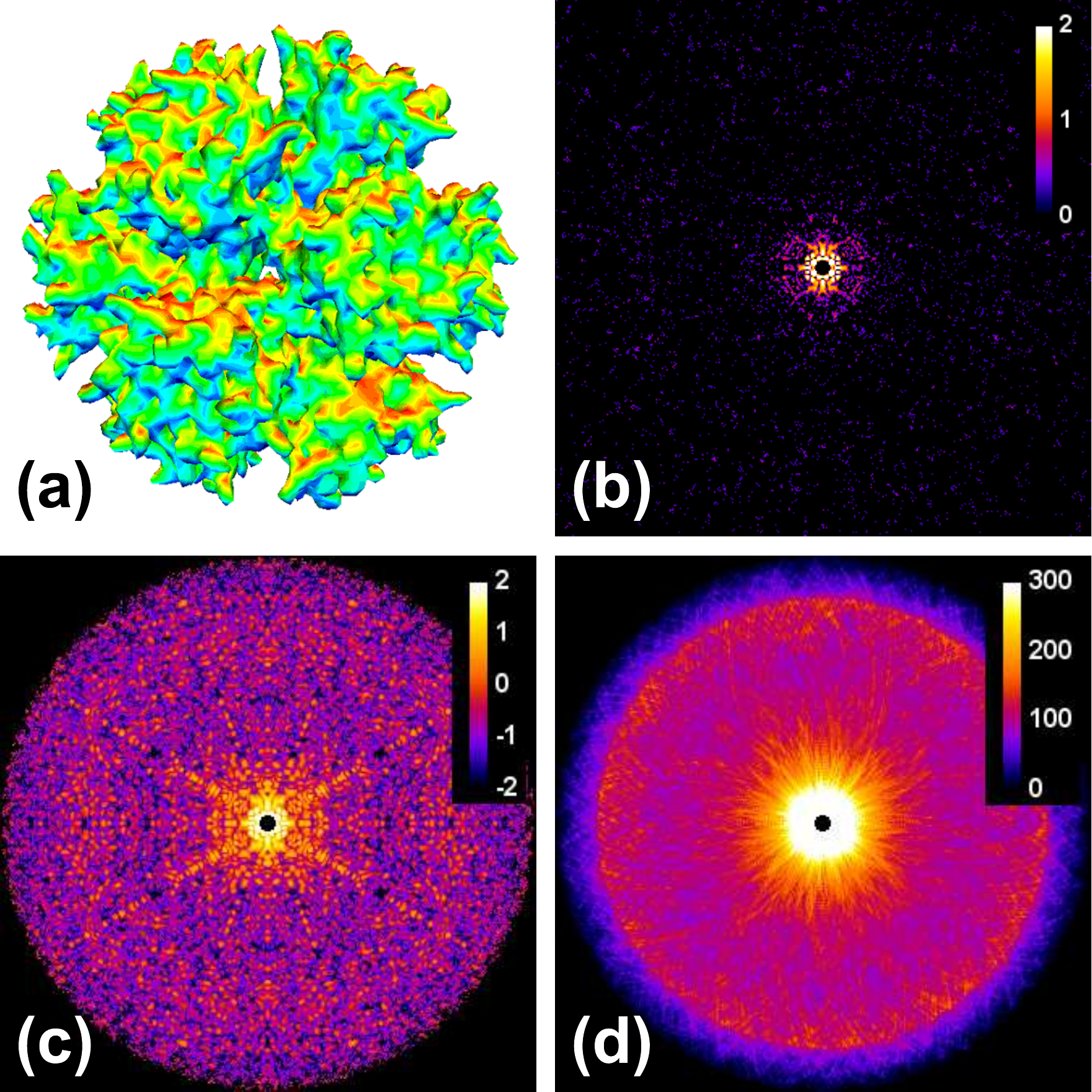}}
\caption{(Color online) Human adenovirus penton base 2 12 chimera 2c6s structure with the icosahedral symmetry. (a),(b),(c) The same as in Fig. \ref{Ruc_result}, (d) number of diffraction patterns contributing to each voxel of (c) (see section \ref{sec:advanced} for details). All diffraction patterns are presented in logarithmic scale.}
\label{2c6s_result}
\end{figure}

The transformation to a Cartesian coordinate system in reciprocal space is performed in the following way. The whole reciprocal space is divided into elementary set of 3D voxels with the size corresponding to the pixel size of the detector. Each voxel could contain few values of the measured intensities that effectively increase signal in the 3D intensity intensity distribution (see Fig. \ref{2c6s_result},d). All intensities in each voxels are averaged and the full 3D dataset is obtained.

Orientation determination problem for the symmetrical structure (2c6s) without introducing the symmetry is more difficult. 
This is due to the fact that instead of one dense spot in Fig. \ref{Plot2Dangles} there will be 60 (for structure with icosahedral symmetry) less dense spots in 3D angular space. So it is harder for algorithm to select the right orientation. The accuracy (or time) of orientations determination can be greatly improved if the symmetry of the sample is known. But we want to underline the important feature of the algorithm, that it can be used even for symmetrical data with unknown symmetry and also for the structures with pseudo-symmetry.

There is one more important issue for speed and accuracy optimization while processing low flux data with noise. For the initial orientations determination there is no need to process high-Q region of diffraction pattern where the radially averaged photon counts is less then approximately 1-2 photons per Shannon pixel (a pixel with sampling rate equal to one). The region with smaller photon counts just lowers the accuracy of orientation determination based on common arcs. Of course this argument cannot be applied to objects with highly anisotropic scattering in different directions, like, for example, pyramids \cite{Yefanov_APL_2009}. At the same time the final 3D reciprocal space and 3D angular refinement can be performed for the full datasets thus the whole procedure does not lower the resolution.

The described in this chapter parameters for the two analyzed structures were: $N_{base} = 64$ base patterns, angular step for each Euler's angle was 3$^\circ$, $A_{tol}=5^\circ$ and $N_{angl} = 30$. The initial orientation determination of the simulated diffraction patterns for both model structures was performed for the circular low-Q region, $150$ pixels in diameter, whereas the 3D refinement for 512 (8RUC) and 360 (2BTV) pixels with angular step of 0.5$^\circ$ in the range of 5$^\circ$ near the previously found position. The initial orientation determination of 23 328 patterns with a 3$^\circ$ angular step and $N_{base} = 64$ base patterns took about one week on a single 8-core computer. The refinement of the found orientation took about one day.







\bibliography{orient}

\newpage

\end{document}